\documentclass{llncs}

\usepackage{cite}

\usepackage{graphicx}
\DeclareGraphicsExtensions{.pdf,.jpeg,.png}

\usepackage{amsmath}
\usepackage{amssymb}
\usepackage{amsfonts}
\usepackage[mathscr]{euscript}

\usepackage{url}

\usepackage{textgreek}
\usepackage{bbding}

\newcommand\DOECopyrightFootnote[1]
{
 \begingroup
 \renewcommand\thefootnote{}\footnote{#1}%
 \addtocounter{footnote}{-1}%
 \endgroup
}

\begin{document}

\title{Pattern-based Modeling of High-Performance Computing Resilience}

\author{
        Saurabh Hukerikar~\Envelope, 
	Christian Engelmann
}
\institute{
        Computer Science and Mathematics Division\\
        Oak Ridge National Laboratory\\
        Oak Ridge, TN, USA\\
        Email: \{hukerikarsr, engelmann\}@ornl.gov
}

\maketitle

\begin{abstract}
With the growing scale and complexity of high-performance computing (HPC) systems, resilience solutions that ensure continuity of service despite frequent errors and component failures must be methodically designed to balance the reliability requirements with the overheads to performance and power.   
Design patterns enable a structured approach to the development of resilience solutions, providing hardware and software designers with the building block elements for the rapid development of novel solutions and for adapting existing technologies for emerging, extreme-scale HPC environments.   
In this paper, we develop analytical models that enable designers to evaluate the reliability and performance characteristics of the design patterns. These models are particularly useful in building a unified framework that analyzes and compares various resilience solutions built using a combination of patterns.

\end{abstract}

\DOECopyrightFootnote{This work was sponsored by the U.S. Department of Energy's Office of Advanced Scientific Computing Research. This manuscript has been authored by UT-Battelle, LLC under Contract No. DE-AC05-00OR22725 with the U.S. Department of Energy. The United States Government retains and the publisher, by accepting the article for publication, acknowledges that the United States Government retains a non-exclusive, paid-up, irrevocable, world-wide license to publish or reproduce the published form of this manuscript, or allow others to do so, for United States Government purposes. The Department of Energy will provide public access to these results of federally sponsored research in accordance with the DOE Public Access Plan (http://energy.gov/downloads/doe-public-access-plan).}

\section{Introduction}
\label{sec:Introduction}
 
Many of the choices that drive hardware and software component designs in emerging extreme-scale high-performance computing (HPC) systems are made to deliver maximum application performance, but are also subject to the constraints of cost, power and reliability. While HPC system architectures have evolved significantly over the past decade, these constraints are expected to force further dramatic changes to the system stack to achieve exascale performance.  
Recent system architectures have emphasized increasing on-chip and node-level parallelism in addition to complex memory architectures consisting of deeper hierarchies and diverse technologies \cite{Shalf:2010}. The software infrastructure, including the system software, middleware and tools, has continued to evolve to keep up with these changes to the system architectures to drive application performance on these extreme-scale computers.  

The reliability and availability of the recent generation of HPC systems have been degrading in comparison to their predecessors \cite{Geist:2016}. This trend is projected to cause future extreme-scale systems to experience unprecedented rates of faults, which will make it difficult to accomplish productive work. The increasingly complex, multicomponent hardware and software environment only makes the challenge of detection of faults in a timely manner, containment of error propagation and mitigation of the impact of error and failure events more difficult.
Resilience solutions must protect the correctness of HPC applications in the presence of faults, errors and failures arising from a multitude of sources, including the system environment, the interactions between platform hardware and system software components and applications, and variability in behavior of hardware components, while seeking to limit the performance and power overhead they impose on the system.  

To navigate the complexities of this emerging landscape of HPC design, we proposed a structured approach to designing HPC resilience solutions based on the concept of design patterns \cite{Hukerikar:2017}. In general, a design pattern is a general reusable solution to a commonly occurring problem within a given context in any design discipline. A pattern provides a description or template for how to solve a problem that may be adapted to specific context. Resilience patterns describe solutions to confront faults and their consequences. The patterns describe techniques for detection, containment and mitigation of faults, errors and failure events. They can be instantiated at any layer of the system stack. The resilience design patterns serve as building block elements for designing complete solutions, and are useful for the exploration of design alternatives for a target HPC system environment and application workload. Section \ref{sec:Background} describes the concept of patterns and summarizes the different types of resilience patterns that are organized in a catalog.  

The development of resilience solutions through composition of various design patterns lends structure to the design and implementation process by compelling designers to consider the key issues of protection coverage, fault model, handling capability, etc. However, objectively selecting pattern solutions that have been examined and utilized successfully in a specific context with the intention of adapting them to a new architecture or software environment of a future system requires criteria based on a quantitative foundation. 
Mathematical models of hardware or software components, or even entire HPC systems, which are solved either analytically or through discrete event simulation, are useful to HPC designers for predicting resilient behavior of the system in the presence of various fault, error and failure events, without having to build the component or system. 
This paper develops models for analytical evaluation of reliability and performance measures of the various resilience design patterns in our pattern catalog. These models are presented in Section \ref{sec:Models}. 
The models are designed to capture the interaction between the resilient behavior and the performance overhead incurred by instantiating a specific pattern. 
Section \ref{sec:Evaluation} discusses approaches to calculate reliability and performance of a solution built by combining several patterns.

\section{Background: Resilience Design Patterns}
\label{sec:Background}

\subsection{Concept}
Design patterns identify the key aspects of a solution to common problem, and presents the solution in the form of an abstract description, which provides designers with guidelines on how to address the problem. Patterns capture the best-known techniques to solve a problem. We developed resilience design patterns \cite{Hukerikar:2017} to support a systematic approach to designing and implementing new resilience solutions and adapting existing solutions to future extreme-scale architectures and software environments. 

The patterns describe the design decisions and trade-offs that must be considered when applying a pattern solution to a specific context. The descriptions encourage designers to reason about the impact of applying a solution on a system's performance scalability and power consumption overhead as well as consider implementation issues. Based on the patterns, we developed a framework that enables designers to comprehensively evaluate the scope of protection domain and the handling efficiency of resilience solutions.

The basic template of a resilience design pattern is defined in an event-driven paradigm, in which each resilience design pattern consists of a {\em behavior} and a set of {\em activation} and {\em response interfaces}. The patterns present solutions to specific problems in detecting, recovering from, or masking a fault, error or failure event. The pattern descriptions are abstract and they may be implemented by HPC applications' algorithms, numerical libraries, system software, or even in the hardware architectures. We have organized the resilience design pattern as a catalog that contains detailed descriptions of the patterns. The catalog is available as a specification document \cite{RDP:Spec}, in which each resilience pattern is presented using a structured format to enable designers to quickly discover whether the pattern solution is suitable to the problem being solved. 

\subsection{Classification}
We developed a pattern classification scheme that organizes the resilience patterns in a layered hierarchy, in which each level addresses a specific aspect of the problem. The classification enables designers to separately reason about the patterns that define the scope of the protection domain and those that define the semantics of the detection, containment and mitigation. The hierarchical organization of the patterns permits system architects to work on the overall organization of the solutions by analyzing the integration of various resilience patterns across the system stack while designers of individual hardware and software components can focus on implementation of the patterns.  

Resilience in the context of HPC systems and its applications has two key dimensions: (1) forward progress of the system; (2) data consistency in the system. Based on these factors, we organize the resilience design patterns into two major categories, \textbf{state} patterns and \textbf{behavioral} patterns.
The behavioral patterns identify detection, containment, or mitigation actions that enable a system to cope with the presence of a fault, error, or failure event. These patterns are organized hierarchically and they include \textbf{strategy}, \textbf{architectural} and \textbf{structural} patterns. 

The strategy patterns define high-level polices of a resilience solution. Their descriptions are deliberately abstract to enable hardware and software architects to reason about the overall organization of the techniques used and their implications on the full system design. These patterns describe the overall structure of the solution and the key attributes of the solution and their capabilities independent of the layer of system stack and hardware/software architectural features. The architectural patterns convey specific methods necessary for the construction of a resilience solution. They explicitly convey the type of fault, error or failure event that they handle and provide detail about the key components and connectors that make up the solution. The structural patterns provide concrete descriptions of the solution rather than high-level strategies. They comprise of instructions that may be implemented in hardware/software components. While the strategy and architectural patterns serve to provide designers with a clear overall framework of a solution and the type of events that it can handle, the structural patterns express the details so they can contribute to the development of complete working solutions.

\subsection{Designing Resilience Solutions using Patterns}
Each pattern in the resilience design pattern catalog presents a solution to a specific problem in detecting, containing or mitigating a fault, error or failure event. In order to construct complete resilience solutions designers must identify patterns that provide each of these capabilities and apply them to a well-defined protection domain. Therefore, a complete solution consists of at least one state pattern (defining scope
of the protection domain), and one or more behavioral patterns (supporting a combination of
detection, containment and mitigation solutions).

For hardware and software designers to make practical use these patterns in the development of resilient versions of their designs, we have developed a design framework that a set of guidelines are necessary to combine the patterns and refine their implementations. The framework is based on design spaces that are arranged
in a hierarchy. By working through the design spaces, designers can convert initial outline of the resilience solution into a concrete implementation by considering the layer of abstraction for the pattern implementation, scalability of the solution, portability to other architectures, dependencies on any hardware/software features, flexibility to adapt the solution to accelerated fault rates, capability to handle other types of fault and error events, the performance and performance overheads.

\section{Reliability and Performance Models for Resilience Design Patterns}
\label{sec:Models}

The models are intended to be useful for predicting the reliability and performance characteristics of solutions built using design patterns in a notional extreme-scale system that may use different plausible architectures and configurations that consist of different node counts, and may use different software environments.  
Therefore, we present the analytic models for the various architecture patterns in our catalog because these patterns explicitly specify the type of event that they handle and convey details about the handling capabilities and the components that make up the solution in a manner independent of the layer of system stack and hardware/software architectural features. For the checkpoint and rollback pattern, we present models for the derivative structural patterns due to their widespread use in HPC environments. The models for the patterns provide a quantitative analysis of the costs and benefits of instantiating specific resilience design patterns. The models may be applied to an individual hardware or software component, which is a sub-system, or to a full system that consists of a collection of nodes capable of running a parallel application. 

Although the future extreme-scale systems may not look at all like the systems of today, we assume that the notional system consists of multiple processing nodes, and that the parallel application partitions the work among tasks that run on these nodes that cooperate via message passing to synchronize. Therefore, we use the following notation in the descriptions of the models: \textit{N}: number of tasks/processes in the parallel application; \textit{M}: total number of messages exchanged between the tasks/processes of the application; \textit{P}: the number of processors in the system; \textit{T$_{system}$}: the operation time of the system, or the execution time of an application.  

In general, we assume that the event (whether fault, error or failure) arrivals follow a Poisson process, the probability of an event is F(t). The reliability of the system is:
\begin{equation}
R(t) = 1 - F(t) 
\label{eq:reliability1}
\end{equation}
which indicates the probability that the system operates correctly for time t.

If the scope of the system captured by the state pattern has an exponential event distribution, the reliability of the system takes the form: 
\begin{equation}
R(t) = 1 - e^{-t/\eta}
\label{eq:reliability2}
\end{equation}
where $\eta$ is the mean time to interrupt of the system, which may be calculated as the inverse of the failure rate of the system.

\subsection{Fault Diagnosis Pattern Model}
The fault diagnosis pattern identifies the presence of the fault and tries to determine its root cause. 
Until a fault has not activated into an error it does not affect the correct operation of the system, and therefore the pattern makes an assessment about the presence of a defect based on observed behavior of one or more system parameters. To incorporate this pattern in an HPC environment requires inclusion of a monitoring component. The pattern uses either effect-cause or cause-effect analysis on the observed parameters of a monitored system to infer the presence of a fault. The performance overhead of this pattern may be expressed as:
 
\begin{equation}
T_{system} = T_{0} + \sum_{k=1}^{n} t_{inference}/\beta
\label{eq:diagnosis1}
\end{equation}

where $n$ is the number of observed parameters of the monitored system and $\beta$ is the frequency of polling the monitored system. Since the pattern only identifies faults, but does not remedy them, there is no tangible improvement in reliability of the system when this pattern is instantiated.  

\subsection{Reconfiguration Pattern Model}
The reconfiguration pattern entails modification of the interconnection between components in a system, such that isolates the component affected by a fault, error or failure event, preventing it from affecting the correct operation of the overall system. The pattern may cause the system to assume one of several valid configurations that are functionally equivalent to the original system configuration, but results in system operation at a degraded performance level. 

To simplify the derivation of the reliability and performance models, we assume that the system consists of n identical components. The performance of the system for the loss of a single component may be expressed as:
\begin{equation}
T_{system} = T_{FF} + (1 - T_{FF}).\frac{n-1}{n} + T_{R}  
\label{eq:reconfig1}
\end{equation} 

where T$_{FF}$ represents the operational time before the occurrence of the event, and T$_{R}$ is the system downtime on account of the delay for reconfiguring the n-1 components. 

The reliability of the system may be expressed as: 
\begin{equation}
R(n,t)  = 1 - \prod_{i=1}^{n}(1-R_{i}(t))
\label{eq:reconfig2}
\end{equation}

This equation assumes that the fault events are independent and are exponentially distributed. 

\subsection{Rollback Recovery Pattern Model}
The checkpoint-recovery architectural pattern is based on the creation of snapshots of the system state and maintenance of these checkpoints on a persistent storage system during the error- or failure-free operation of the system. Upon detection of an error or a failure, the checkpoints/logged events are used to recreate last known error- or failure-free state of the system, after which the system operation is restarted. The rollback recovery pattern is a derivative of the checkpoint-recovery provides rollback recovery, i.e., based on a temporal view of the system's progress, the system state recreated during the recovery process is a previous correct version of the state of the system.  

The pattern requires interruption of the system during error or failure-free operation to record the checkpoint, which incurs an overhead. Therefore, the operational lifetime of the system can be partitioned into distinct phases, which include the regular execution phase (\textit{o}), the interval for creating checkpoints ($\delta$), and the interval for recovery upon occurrence of an event ($\gamma$) to account for the operational state lost on account of the event.

The performance of the system in absence of any error or failure events may be expressed as:
\begin{equation}
T_{system} = o + \delta/r
\label{eq:cr1}
\end{equation}

where \textit{r} is the rate of checkpointing.

The performance of the system in the presence of failure events, assuming an exponential event rate of e$^{-t/\eta}$ ($\eta$ is the mean time to interrupt of the system) may be modeled as: 
\begin{equation}
T_{system} = (T_{FF} + \gamma)/\eta 
\label{eq:cr2}
\end{equation}

where T$_{FF}$ = o + $\delta$/r

The reliability of a system using the rollback recovery pattern may be modeled as: 
\begin{equation}
R(t) = 1 -  e^{-(T_{FF} + \gamma)/\eta} 
\label{eq:cr3}
\end{equation}

for systems in which an event occurs before the interval T$_{FF}$ + $\gamma$, and $\eta$ is the mean time to interrupt.

\subsection{Roll-Forward Recovery Pattern Model}
The roll-forward pattern is a structural pattern, which is also a derivative of the checkpoint recovery pattern. It uses either checkpointing or log-based protocols to record system progress during error- or failure-free operation. The recovery entails the use of checkpointed state and/or logging information to recreate a stable version of the system identical to the one right before the error or failure occurred. The roll-forward pattern may also use online recovery protocols that use inference methods to recreate state.

The roll-forward pattern also requires the system to record system and/or message state during fault-free operation. The system performance may be calculated using: 
\begin{equation}
T_{system} = o + \delta/r
\label{eq:rf1}
\end{equation}

where \textit{r} is the rate of checkpointing or message logging.

The performance of the system in the presence of failure events may be captured using: 
\begin{equation}
T_{system} = (T_{FF} + \gamma)/\eta 
\label{eq:rf2}
\end{equation}

where T$_{FF}$ = o + $\delta$/r.

When the roll-forward pattern instantiation uses message logging, the term $\delta$ in these equations is calculated as the logging interval: $\delta$ = M . t$_{logging}$.

The reliability of the system that uses the rollforward pattern capability may be modeled as:
\begin{equation}
\begin{split}
R(t) & = 1 -  e^{-(T_{FF} + M.t_{logging})/\eta} \text{[for message logging implementations]} \\
     & = 1 -  e^{-(T_{FF} + \gamma)/\eta}        \text{[for checkpointing implementations]} 
\end{split}
\label{eq:rf3}
\end{equation}

assuming an exponential event arrival and $\eta$ is the mean time to interrupt of the system. 

\subsection{Redundancy Pattern Model}
The redundancy pattern is based on a strategy of compensation since it entails creation of a group of N replicas of a system. The replicated versions of the system are used in various configurations to compensate for errors or failures in one of the system replicas, including fail-over, active comparison for error detection, or majority voting for detection and correction by excluding the replica whose outputs fall outside the majority. The use of the redundancy pattern incurs overhead to the system operation independent of whether an error or failure event occurs. 

For parallel application, the overhead depends on the scope of replication, which may include aspects such as the amount of computation performed by the tasks, the communication between them, etc. The overhead also depends on factors such as the degree of redundancy, placement of the replicas on the system resources. Therefore, to develop a precise mathematical model that represents each of these factors is complex. To simplify the analysis, we partition the operation time of the system into the ratio of the time spent on the redundant operation $\mathcal{A}$ and the time . This partitioning can be logically defined by the scope of the state patterns; (1 - $\mathcal{A}$) is the fraction outside the scope of the state pattern, for which no redundancy is applied. Since the term \textit{t} is taken as the base execution time of the application, the time $\mathcal{A}$.t is the time of system operation for which redundancy is applied, while (1 - $\mathcal{A}$).t is the remaining time. The term \textit{d} refers to the degree of redundancy, i.e., the number of copies of the pattern behavior that are replicated.

\begin{equation}
T_{system} = T_{S}.((1 - \mathcal{A}) + \beta.\mathcal{A}))  + T_{MV}
\label{eq:redundancy1}
\end{equation}

where $\beta$ is 1 when the state pattern is replicated in a space redundant manner and is equal to \textit{d} when applied in a time redundant manner. The term T$_{S}$ is serial operation time of the system and the term T$_{MV}$ represents the time spent by the majority voting logic to detect output mismatches.  

Assuming the mean time to interrupt of the system that uses the redundancy pattern is $\lambda$, then the reliability of the system may expressed as:
\begin{equation}
R(t) = 1 - \prod_{i=1}^{d} t/\lambda = 1 - (t/\lambda)^{d}  
\label{eq:redundancy2}
\end{equation}

\subsection{Design Diversity Pattern Model}
When a design bug exists in a system design or configuration, an error or failure during system operation is often unavoidable. Therefore, the detection and mitigation of the impact of such errors or failures is critical. The n-version design pattern applies distinct implementations of the same design specification created by different individuals or teams. The N versions of the system are operated simultaneously with a majority voting logic is used to compare the results produced by each design version. Due the low likelihood that different individuals or teams make identical errors in their respective implementations, the pattern enables compensating for errors or failures caused by a bug in any one implementation version.

Assuming that there are n versions of the system scope encapsulated by the state pattern, 1 $\geq$ i $\leq$ n, then the probability that only version i executes its function correctly while the remaining versions produce an incorrect outcome:
\begin{equation}
P(A) = \sum_{k=1}^{n+1} P(A_{k}) 
\label{eq:nversion1}
\end{equation}

where the P(A$_{k}$) is the probability that only the version A$_{k}$ out of the n versions produces the correct outcome, while the remaining versions produce an incorrect outcome.

The probability density function (PDF) describing the probability of failure occurring during the system operation may be expressed as:
\begin{equation}
P(t) = ( (1 - P(V))\sum_{k=1}^{n+1} P(A_{k})  + P(V) )
\label{eq:nversion2}
\end{equation}

where the P(V) represents the probability that the majority voting procedure cannot select the correct result from at least two correct versions. Therefore, the reliability of the system using the n-version design at time t may be calculated in terms of this probability:

\begin{equation}
R(t) = 1 - ( (1 - P(V))\sum_{k=1}^{n+1} P(A_{k})  + P(V) ). F(t)
\label{eq:nversion3}
\end{equation}

where F(t) = e$^{-t/\eta}$ is the failure rate assuming exponential event arrival rate.

\section{Model-based Evaluation of Resilience}
\label{sec:Evaluation}

The design of complete resilience solutions often requires the composition of multiple resilience design patterns. In a complex HPC environment with numerous hardware and software pattern instantiations in the various components, the resilience to different fault events is managed by this well-defined system of patterns. To developed a combined evaluation of the reliability and performance characteristics of a real system that consists of several pattern solutions implemented across the system stack requires composition of the pattern models. 

For the simplified case of a system configuration that consists of N independent components or tasks such that, if any one of the system components or tasks fails, the entire system fails, the overall reliability of the system may be modeled as: 

\begin{equation}
R_{system} = R_{1} \times R_{2} \times R_{3} \times . . .  R_{N}  
\label{eq:sys-reliability}
\end{equation}
where the reliability R$_{i}$ of a component is a function of the resilience pattern that it instantiates. For such a configuration, the performance overhead of applying patterns to the N components in the system is additive. 

For more intricate analytic evaluation of the performance and reliability, more complex models must be developed. There are several paradigms that are useful for this purpose, including fault trees, block diagrams, reliability \& task graphs, Markov \& semi-Markov chains, stochastic Petrinets, etc. 
Analytical models that use Markov models are useful to model the intricate dependencies between the pattern solutions in a complex multicomponent HPC environment. Markov chains are state-space-based methods that consist of states representing various conditions associated with the system, and the transition between states, which represent the changes in system state or configuration due to the occurrence of a simple or compound event such as the malfunction or failure of one or more components in the system. The assessment of system resilience using Markov models for a multicomponent HPC environment that experiences different modes of faults, as well as a model for the combined evaluation of performance and reliability is the subject of ongoing research.   

\section{Related Work}
\label{sec:RelatedWork}

Much research has been done on modeling techniques and tools that are useful for reliability and performance analysis  of various computing systems and applications. These approaches may broadly be categorized into \cite{Pham:2006}: (i) \textit{structural modeling}, which highlights the relationships between the system components using representations such as block diagrams, reliability graphs and fault trees. These models assume stochastic independence between system components; (ii) \textit{state-space models}, which model the dependencies among system components and use representations such as Markov chains. These models are significantly more complex due to the need for as many as 2$^{n}$ states in the Markov representation for n system components; and, (iii) \textit{hierarchical models}, which balance the speed of analysis with the accuracy of the model by combining abstract structural models with the detailed Markov models \cite{Geist:1990}. There have also been several advances in performability analysis, which aim to model the interaction between failure recovery behavior and performance in a composite manner \cite{Beaudry:1978} \cite{Trivedi:1993}.  

Due to the dominance of checkpoint and rollback approaches in high-performance computing systems, several approaches have been proposed for calculating the reliability and performance measures of systems that use this solution.    
Analytic models have been developed for determining the optimum intervals for checkpoints \cite{Young:1974}. For applying such analysis to large-scale cluster-based HPC systems, the model has been adapted to meet the goal of minimizing the overall application run time \cite{Daly:2006}. For understanding the viability of rollback recovery on extreme-scale systems, models for prediction of its performance have been proposed. This model was developed to evaluate the rollback recovery solution for petascale HPC systems \cite{Elnozahy:2004}. An optimal checkpoint and rollback model has been devised that incorporates a reliability function obtained from the analysis of historical failure data from the system event log files \cite{Liu:2007}. There have also been efforts to develop models that analyze the combination of checkpoint restart with redundancy techniques for MPI applications \cite{Elliott:2012}, and for multilevel checkpointing solutions \cite{Di:2014}.

\section{Conclusion}
\label{sec:Conclusion}

For future extreme-scale HPC systems, designers will be required to work within tight constraints of cost, power and reliability to achieve greater performance. The use of design patterns enables the exploration of alternative solutions for a specific context and provides a framework to combine individual patterns into complete solutions.  
The performance and reliability models for resilience design patterns presented in this paper allow us to develop measures to analyze the solutions built using the patterns. 
While these models for the architecture patterns are not detailed enough and refining them will lead to considerable added complexity, they predict the implications of selecting specific combination of patterns for the reliability and performance of a system in a given context.   
The models developed in this paper are designed to be useful for simulation frameworks to examine the effectiveness of a resilience solution for specific fault models and fault rates, as well as to measure the performance and reliability characteristics of the pattern-based solution for different system architectures, software environments and application workloads.

\subsubsection*{Acknowledgements}
This material is based upon work supported by the U.S. Department of Energy, Office of Science, Office of Advanced Scientific Computing Research, program manager Lucy Nowell, under contract number DE-AC05-00OR22725. 

\bibliographystyle{splncs03}
\bibliography{references}

\end{document}